\documentclass{article}
\usepackage[utf8]{inputenc}
\usepackage[margin=1in]{geometry}
\usepackage{amsmath}
\usepackage{amssymb}
\usepackage{amsthm}
\usepackage{mathtools}
\usepackage{xcolor}

\usepackage[pdfencoding=auto, psdextra]{hyperref}
\usepackage[english]{babel}

\newtheorem*{lemma}{Fact}
\newtheorem*{theorem}{Theorem}
\newtheorem*{cor}{Corollary}

\title{The Complexity of Checking Partial Total Positivity\footnote{This work was supported by the 2018 National Science Foundation Grant DMS \#0751964.}}
\author{Daniel Carter\thanks{Department of Mathematics, Princeton University, Princeton, NJ 08544}\and Charles Johnson\thanks{Department of Mathematics, College of William and Mary, Williamsburg, VA 23187-8795 (crjohn@wm.edu)}}
\date{2020-2021}

\begin{document}

\maketitle

\begin{abstract}
    We prove that checking if a partial matrix is partial totally positive is co-NP-complete. This contrasts with checking a conventional matrix for total positivity, for which we provide a cubic time algorithm. Checking partial sign regularity with any signature, including partial total nonnegativity, is also co-NP-complete. Finally, we prove that checking partial total positivity in a partial matrix with logarithmically many unspecified entries may be done in polynomial time.
\end{abstract}

\section{Background}

An $m$-by-$n$ matrix $A$ is \textit{totally positive} (TP) or \textit{totally nonnegative} (TN) if all of its minors (determinants of square submatrices) are positive or nonnegative, respectively. Na\"ively, a matrix has exponentially many minors that must be checked to verify it is TP or TN. However, due to Fekete's Theorem, only the \textit{contiguous} minors, that is, minors whose row and column index sets are consecutive, must be checked to verify total positivity. Even better, only the \textit{initial minors} (contiguous minors whose row or column set includes 1) need to be checked \cite{totallynonnegative}. The number of initial minors in an $m$-by-$n$ matrix is precisely $mn$, and calculating determinants may be done in $O(n^{2.373})$ time, so this means there is a polynomial time algorithm for checking total positivity \cite{detcomplexitynew}. In Section~\ref{dodgson}, we demonstrate a simple cubic-time algorithm to check total positivity based on Dodgson condensation.

A \textit{partial matrix} is one in which some entries are unspecified and free to be chosen. A \textit{completion} of a partial matrix is a choice of values for the unspecified entries, resulting in a conventional matrix of the same size. The \textit{TP completion problem} asks for which partial matrices there exists a completion that is TP \cite{totallynonnegative}. A partial matrix is \textit{partial TP} if all of its minors that consist only of specified entries are positive. Clearly a necessary condition for there to be a TP completion of a partial matrix is that the partial matrix is partial TP. This is not sufficient in general. In Section~\ref{main}, we prove that determining if a partial matrix is partial TP lies in the complexity class co-NP-complete, in stark contrast to the case for conventional matrices. We actually prove a more general result for strongly and weakly sign regular matrices that includes checking partial TP and TN as special cases.

Finally, in Section~\ref{algs}, we provide an algorithm for checking partial TP which runs in exponential time in the number of unspecified entries and polynomial time in the size of the partial matrix. This generalizes to any matrix property which is inherited by all submatrices.

Our principal observation about the complexity of checking partial TP is clearly relevant to the solution of TP completion problems.

\section{TP in Cubic Time}
\label{dodgson}

Dodgson condensation is a method for computing the determinant of a matrix by repeatedly applying the Desnanot-Jacobi-Sylvester identity
\[ \det(M)=\frac{\det(M_1^1)\det(M_n^n)-\det(M_1^n)\det(M_n^1)}{\det(M_{1,n}^{1,n})}, \]
in which $M$ is an $n$-by-$n$ matrix and $M_{a_1,a_2,...}^{b_1,b_2,...}$ denotes the submatrix obtained by deleting rows $a_1,a_2,...$ and columns $b_1,b_2,...$. Dodgson condensation calculates the determinants of ever-larger contiguous minors until the determinant of the whole matrix is found, as follows:
\begin{enumerate}
    \item Set $A\gets M$.
    \item Let matrix $B$ be $(n-1)$-by-$(n-1)$ with entries $b_{i,j}=a_{i,j}a_{i+1,j+1}-a_{i,j+1}a_{i+1,j}$. Evidently, the entries of $B$ are the contiguous $2$-by-$2$ minors of $M$.
    \item Let matrix $C$ be $(n-2)$-by-$(n-2)$ with entries $c_{i,j}=(b_{i,j}b_{i+1,j+1}-b_{i,j+1}b_{i+1,j})/a_{i+1,j+1}$. In light of the identity mentioned above, the entries of $C$ are the contiguous $3$-by-$3$ minors of $M$.
    \item Set $A\gets B$ and $B\gets C$, and repeat step 3 to obtain a new matrix $C$ of dimension one less than before. This matrix consists of the contiguous minors of $M$ one size greater than before. Do this until the resulting matrix is $1$-by-$1$. Its entry is the determinant of the original matrix $M$.
\end{enumerate}

If ever an interior entry of the previous matrix is zero, step 2 is ill-defined. However, note that we compute every contiguous minor of the matrix at some point in the algorithm, so if any entry of any of the matrices is nonpositive, we can immediately conclude the original matrix was not TP. If instead the above algorithm terminates with all entries encountered positive, then all contiguous minors of the original matrix were positive, and it is TP. It is clear that we may apply the same algorithm on non-square matrices, stopping when we obtain a matrix with either only one row or only one column.

Analysis of the complexity of this algorithm is straightforward. To get the contiguous $(k+1)$-by-$(k+1)$ minors from the contiguous $k$-by-$k$ minors, this algorithm performs two multiplications, one subtraction, and one division per $(k+1)$-by-$(k+1)$ minor. If the matrix is $m$-by-$n$ with $m\le n$, there are $(m-k)(n-k)$ such minors, and the largest minors are $m$-by-$m$ (so the largest value of $k$ is $m-1$). Therefore the asymptotic runtime is given by
\[ \sum_{k=1}^{m-1}(m-k)(n-k)c, \]
with $c$ the cost of performing the arithmetic operations. This sum is $(3m^2n-m^3-3mn+m)\cdot c/6$, which is $O(m^2n)$. For square matrices this is cubic time.

Note there is no general analogue to Fekete's Theorem for total nonnegativity, so the cubic time algorithm presented here cannot be straightforwardly adapted to get a cubic time algorithm for determining TN. There are non-TN matrices where all contiguous minors have nonnegative determinant, such as the one below:
\[ \begin{bmatrix}
1 & 0 & 1 \\
1 & 0 & 0
\end{bmatrix}. \]

\section{Partial TP is Co-NP-complete}
\label{main}

A matrix is called \textit{weakly sign regular} (WSR) with signature $(e_i)$, $e_i\in\{-1,1\}$ for all $i$, if all of its $k$-by-$k$ minors $A$ satisfy $e_kA \ge 0$ for all $k$. If the minors satisfy $e_kA > 0$ for all $k$, the matrix is \textit{strictly sign regular} (SSR). Partial matrices are partial WSR if their $k$-by-$k$ specified minors satisfy $e_kA \ge 0$ for all $k$ and similarly for partial SSR. Note that TP (TN) is the special case of SSR (WSR) with $e_i=1$ for all $i$.

A \textit{biclique} is an induced subgraph of a bipartite graph that is a complete bipartite graph; a biclique is \textit{balanced} if it has the same number of vertices in both parts. The \textsc{balanced biclique} problem asks if there is a balanced biclique in a given bipartite graph with at least $2k$ for given integer $k$. The \textsc{balanced biclique} problem is shown to be NP-complete in \cite{npcomplete} by a transformation from the \textsc{clique} problem.

We reduce the \textsc{balanced biclique} problem to instances of the complement of \textsc{partial $(e_i)$-SSR} and \textsc{partial $(e_i)$-WSR}, the problems of checking if a given partial matrix is partial SSR/WSR with a fixed signature $(e_i)$. This shows that \textsc{partial $(e_i)$-SSR} and \textsc{partial $(e_i)$-WSR} are co-NP-hard. It is easy to show they are also co-NP, so they are co-NP-complete. As special cases, both \textsc{partial TN} and \textsc{partial TP} are also co-NP-complete.

First, we recall a known fact that may be found in \cite{ando} and elsewhere. It may be proven inductively by ``bordering'' \cite{totallynonnegative}.

\begin{lemma}
\label{lem}
For any positive integers $m$ and $n$ and signature $(e_i)$, there exists an $m$-by-$n$ SSR (WSR) matrix with signature $(e_i)$.
\end{lemma}

We then have our main theorem:

\begin{theorem}
Both \textsc{partial $(e_i)$-SSR} and \textsc{partial $(e_i)$-WSR} are co-NP-complete for any signature $(e_i)$.
\begin{proof}
\textbf{Co-NP-hard:} Given a bipartite graph $G$ with parts $U=\{u_1,u_2,\dots,u_m\}$ and $V=\{v_1,v_2,\dots,v_n\}$ and an integer $k$, we construct a particular $m$-by-$n$ partial matrix $M$. First, let $X$ be an $m$-by-$n$ SSR matrix with signature $(f_i)=(e_1,e_2,\dots, e_{k-1},-e_k,-e_{k+1},\dots)$. If $\{v_i,u_j\}$ is an edge of $G$, let $m_{ij}$ be specified and equal to $x_{ij}$; otherwise let $m_{ij}$ be unspecified.

Now $G$ contains a balanced biclique of size $k$ or greater if and only if $M$ has a $k$-by-$k$ specified minor $A$. This minor, if it exists, satisfies $f_kA>0$. This means $M$ is \textit{not} partial SSR (or WSR) with signature $(e_i)$, since it has a $k$-by-$k$ minor with $-e_kA\not>0$ ($-e_kA\not\ge0$). Thus, \textsc{partial $(e_i)$-SSR (WSR)} is co-NP-hard.

\textbf{Co-NP:} Given a partial matrix $M$ which is not partial SSR (WSR) with a given signature $(e_i)$, one can provide a witness $k$-by-$k$ minor $A$ that can be computed in polynomial time which does not satisfy $e_kA>0$ ($e_kA\ge0$), so \textsc{partial $(e_i)$-SSR (WSR)} is co-NP.

We conclude \textsc{partial $(e_i)$-SSR (WSR)} is co-NP-complete.
\end{proof}
\end{theorem}

Of interest is a special case of this theorem:

\begin{cor}
Both \textsc{partial TP} and \textsc{partial TN} are co-NP-complete.
\begin{proof}
The \textsc{partial TP} problem is precisely the \textsc{partial $(1,1,\dots)$-SSR} problem, and the \textsc{partial TN} problem is precisely the \textsc{partial $(1,1,\dots)$-WSR} problem.
\end{proof}
\end{cor}

\section{Algorithms for Checking Partial TP}
\label{algs}

Consider a partial matrix with one unspecified entry at the $(i,j)$ position. Any submatrix with only specified entries that includes row $i$ cannot include column $j$ due to this unspecified entry, and similarly if it includes column $j$ then it cannot include row $i$. This means to check if this partial matrix is partial TP, it is sufficient to check if the two conventional matrices formed by removing row $i$ or column $j$ are TP. These matrices are denoted $M_i$ and $M^j$, following the notation in Section~\ref{dodgson}.

This generalizes to the case of general partial matrices:

\begin{theorem}
If a partial matrix $M$ has an unspecified entry at the $(i,j)$ position, then it is partial TP if and only if the two partial matrices $M_i$ and $M^j$ are partial TP.
\begin{proof}
Any fully specified square submatrix of $M$ must lie entirely within $M_i$ or $M^j$ (or both), since it cannot include the $(i,j)$ entry. Hence if both of these partial matrices are partial TP, any specified square submatrix of $M$ has positive determinant, so $M$ is partial TP. The converse is trivial.
\end{proof}
\end{theorem}

This immediately gives a recursive algorithm to check partial TP:
\begin{enumerate}
    \item If the partial matrix $M$ has no unspecified entries, then check if it is TP using the cubic time algorithm presented previously.
    \item Otherwise, say the $(i,j)$ entry is unspecified. Recursively check $M_i$ or $M^j$ for partial TP; the original matrix is partial TP if and only if both submatrices are partial TP.
\end{enumerate}

It is clear this algorithm runs in time $O(2^xn^3)$, where $x$ is the number of unspecified entries in the partial matrix, since each unspecified entry generates at most two subproblems with at most $x-1$ unspecified entries, for a total of at most $2^x$ subproblems. Consequently, if $x\le c\log n$ for a fixed $c$, this algorithm runs in polynomial time, with the degree depending on $c$. It is also worth noting that the number of specified minors of an $m$-by-$n$ partial matrix is at most $\sum_{k=1}^{\min(m,n)}\binom{m}{k}\binom{n}{k}=\binom{m+n}{n}-1\le 2^{m+n}$, so this algorithm is only faster than brute-force checking every specified minor in the case that there are relatively few unspecified entries.

It is also clear that this generalizes to the case of checking partial SSR or partial WSR for any sign signature, including partial TN, or indeed to any matrix property that is inherited by all submatrices. If one has an algorithm to check a matrix for such a property $P$ that runs in time $O(f(n))$, then the appropriate modification of the above algorithm will check partial matrices for ``partial $P$'' in time $O(2^xf(n))$.

Furthermore, one can connect this result to the bipartite graph construction of the previous section. A \textit{maximal biclique} is a bliclique that is not a subgraph of any larger bliclique. Then we have the following more general theorem:

\begin{theorem}
Let $G$ be the bipartite graph corresponding to a partial matrix $M$. Then $M$ is partial TP iff each of the submatrices corresponding to the maximal blicliques of $G$ are TP.

\begin{proof}
Each fully specified square submatrix of $M$ corresponds to a balanced biclique in $G$. Each balanced biclique is either maximal or a subgraph of a larger biclique, so if all of the maximal bicliques correspond to TP submatrices of $M$, then every specified minor is positive, by virtue of being a minor of a TP submatrix. The converse is trivial.
\end{proof}
\end{theorem}

Therefore any algorithm to enumerate maximal bicliques of a bipartite graph translates into an algorithm for efficiently checking partial TP, simply by checking for TP in each maximal biclique (and it will also inherit a cubic factor from this check).

This is a generalization of the algorithm presented earlier in this section because that algorithm may be obtained from the observation that, if vertices $a_i$ and $b_j$ in bipartite graph $G$ do not have an edge between them, the set of maximal bicliques of $G$ is the union of the sets of maximal bicliques in $G-a_i$ and $G-b_j$.

\bibliographystyle{alpha}
\bibliography{biblio}

\end{document}